\documentstyle[preprint,eqsecnum,aps,epsf]{revtex}
\begin{document}
\tolerance=10000
\hfuzz=5 pt
\baselineskip=24 pt
\draft
\preprint{HD-THEP-97-08 ~~~ JLAB-THY-97-17}

\title{\bf From the Feynman--Schwinger representation to the 
non-perturbative relativistic bound state interaction}
\author{
Nora Brambilla $^{~*}$ 
and Antonio Vairo \thanks{Alexander von Humboldt Fellow} }
\address{\it Institut f\"ur Theoretische Physik, Universit\"at Heidelberg\\ 
Philosophenweg 16, D-69120 Heidelberg, FRG \\ and\\ 
Jefferson Laboratory \\
12000 Jefferson Ave., Newport News, VA 23606, USA \\ and\\
Nuclear/High Energy Physics (NuHep) Research Center, \\
Hampton University, Hampton, VA 23668, USA }
\maketitle

\begin{abstract}
\baselineskip=20 pt 
\noindent 
We write the 4-point Green function in QCD in the Feynman--Schwinger 
representation and show that all the dynamical information are contained 
in the Wilson loop average. We work out the QED case in order to obtain 
the usual Bethe--Salpeter kernel. Finally we discuss the QCD case 
in the non-perturbative regime giving some insight in the nature 
of the interaction kernel.
\end{abstract}


\vfill
\eject

\section{Introduction}

In the context of the study of QCD  bound states via analytic methods 
a lot of interest has been devoted in the last ten years to the so-called 
Feynman--Schwinger formalism \cite{pes83}-\cite{bra96}. 
The main feature of the formalism is that it allows to write the 
4-point Green function (at least in quenched approximation) 
only in terms of a quantomechanical path integral over the quark 
trajectories times a functional depending on the average over 
the gauge fields of the Wilson loop defined by the quark paths. 
Moreover this functional can be expressed in terms of path 
derivatives of the averaged Wilson loop \cite{si,BCP}.   
Once we assume an analytic behaviour for the Wilson loop, which up to now 
can be only given as an external input more or less {\it motivated} by QCD but 
not completely {\it derived} from QCD, the advantages of such a formulation 
are apparent. It permits numerical calculations \cite{nitj} which  
under some conditions can provide results faster and cheaper than 
a traditional lattice calculation. Moreover it allows analytic 
estimates of physical interesting quantities. 
In \cite{BMP,si,BCP,bakbra,bra97} it was possible in this way  to obtain 
the complete heavy quark potential up to the order $1/m^2$ for different 
Wilson loop assumptions and reproduce also the spin-dependent
contributions of Eichten and Feinberg \cite{eichfein} in the appropriate 
limits \cite{bra97}. We can say that the complete semirelativistic 
heavy quark-antiquark dynamics (at least in the form of the interaction 
potential) could be accessed only using this Feynman--Schwinger and path
integral formalism.

On the other side the derivation of the relativistic quark-antiquark 
interaction is a long standing and important problem (see \cite{report} 
for some report papers). All the calculations of phenomenologically relevant 
quantities such as the masses and the form factors for the light hadrons 
rely on the understanding of relativistic quark dynamics
\footnote{In this paper we take into account only analytic models of
the quark dynamics. Of course QCD lattice calculations are an alternative 
and complementary approach to the problem.} .
In the last years a lot of effort has gone into the development  
of light cone Hamiltonians on one side or Bethe--Salpeter-like
and/or Schwinger--Dyson equations on the other.
Some criticism has been made to the latter approach essentially related 
to the loss of gauge invariance \cite{sicon}.
Indeed the manifestly gauge invariance of a physical state is a
relevant concept  when dealing with non-perturbative QCD dynamics.
It is clear that the propagator of a coloured object could not be considered
separately from the other coloured partners since it is connected by a string
to it and this confinement dynamics dominates at large distances.
Another way to put the thing is to say that non-perturbatively the background 
fields and their effect on the quark dynamics are important. Nevertheless 
we believe that when the gauge-invariance issue is properly addressed 
(i. e. the average on all the vacuum fields in the amplitude is correctly 
handled) the resulting effective interaction can be still treated 
in the framework of the Bethe--Salpeter equation and this supplies 
us with a formidable tool for the (numerical) evaluation of a huge number 
of  physical quantities \footnote{For an example of the application of 
the Bethe--Salpeter equation to several phenomenological quantities 
see e.g. \cite{buck}.}.

One of the usually claimed limitations of the Bethe--Salpeter approach is that 
the confining part of the kernel is not known. In the literature 
it is widespreadly used a kernel made of a one-gluon ladder short range 
part plus a long range confining part suggested by a trivial 
relativistic generalization of the  static linear potential. 
This amounts to consider a kernel depending only on the momentum transfer 
$Q$ and with the form $1/ Q^4$. The Lorentz structure of the confining 
kernel is suggested to be a scalar again on the basis of the potential 
(which actually pertains to a complete different dynamical region, 
we point out) or simply phenomenologically treated as a vector which is 
chirally symmetric. However, all these assumptions run into great conceptual 
and concrete difficulties and it emerges that the kernel should be 
more complicate that a pure convolution  type \cite{dif}.
Another approach deals with Bethe--Salpeter and Schwinger--Dyson ``coupled'' 
equations with a kernel inspired by the lattice evaluation 
of the gluon propagator in a given gauge\footnote{In this approach 
the main features of the  phenomenology  connected with the chiral symmetry
can be qualitatively  reproduced  using a generic infrared enhanced 
gluon propagator. This is a further motivation of our believe 
that the characteristics of the light mesons can be well understood in a
Bethe--Salpeter framework.}\cite{roberts}.

The main motivation of this paper is to investigate the nature 
of the fully relativistic quark-antiquark dynamics in the form of 
a Bethe--Salpeter kernel working with the Feynman--Schwinger representation 
of the quark-antiquark gauge-invariant Green function \footnote{
Several attempts have been made also recently to obtain in the 
Feynman--Schwinger formalism a Bethe--Salpeter kernel for the QCD 
bound state \cite{bra96}, but the problem is still open.}. 
The idea is to use this representation, that displays the complete dynamics 
factorized in the Wilson loop, in order to enforce the information 
we have on the Wilson loop behaviour directly on the Bethe--Salpeter 
kernel by means of a completely relativistic and non-perturbative 
procedure. This means that, starting with a form for the Wilson loop
we are able to establish the leading  Feynman graphs that make up the 
interaction kernel. Moreover if we  use a Wilson loop behaviour 
containing the relevant part of the confining dynamics we will end up 
with the relevant part of the confining kernel. As it will become clear, 
the task is not simple in the case of quarks with spin.

In practice a good part of the paper is devoted to the technical 
setting up of the formalism.  As an application we derive 
the leading binding contribution to the Bethe--Salpeter kernel 
in QED (the one photon exchange graph). This supplies us with a 
technique and a definite language to apply in QCD. The up to now available 
assumptions on the behaviour of the Wilson loop average seem not to 
allow easy extensions. Therefore, we suggest the use of the Fock--Schwinger 
gauge in order to implement (as in the QCD sum rules approach) 
non-perturbative physics in the Wilson loop, leaving 
the structure of the Wilson loop average  as close 
as possible to the QED one. We indicate the graphs relevant 
to the quark--antiquark binding that make up the interaction kernel. 

The paper has the following structure. In section 2 and section 3 
we derive  the 2-point and 4-point Green functions in the Feynman--Schwinger 
formalism. In section 4 we apply the formalism to QED and in section 5 
we discuss QCD  and draw some conclusions.  

\section{The Feynman--Schwinger repre\-sen\-ta\-tion 
of the fer\-mion propagator}

The aim of this section is to represent in terms of a quantomechanical path 
integral the fermion propagator $S$ of a particle $m$ in an external gauge 
field $A$. We assume $A$ to be the non-Abelian gauge field associated  
with the gluon in QCD. Therefore where necessary we explicitly denote 
with the symbol ${\rm P}$ the path ordering prescription. 
In the Abelian case this prescription is obviously not needed. 

$S$ is defined as
\footnote{
All the formula here and in the following are given in the usual Minkowski 
metric  (for our purposes we do not need to introduce 
the Euclidean metric which, of course, would be necessary 
in a more formal discussion).}:
\begin{equation}
S_{\alpha\beta}(x,y;A) \equiv {1\over i}\langle {\rm T} \psi_\alpha(x) 
\bar\psi_\beta(y) \rangle_{A}
\,,
\end{equation}
where the brackets $\langle ~~ \rangle_A$ stand for  the average over 
the fermionic fields in the presence of the external source $A$. 
 $S$ satisfies the equations:
\begin{eqnarray} 
(i\,{D\!\!\!\!/}_x - m)\,S(x,y;A) &=& \delta^4(x-y)\,, \label{Seq}\\
S(x,y;A)(i\,\buildrel{\leftarrow}\over{{D\!\!\!\!/}}_y + m) &=& 
-\delta^4(x-y) \,, 
\label{Seq2}
\end{eqnarray}
where $D^\mu_x \equiv \partial^\mu_x - i\,g\,A^\mu(x)$ and 
$\buildrel{\leftarrow}\over{D}^\mu_x \equiv 
\buildrel{\leftarrow}\over{\partial}^\mu_x + i\,g\,A^\mu(x)$.

If we define 
\begin{equation}
(i\,{D\!\!\!\!/}_x + m)~\Delta(x,y;A) \equiv S(x,y;A)\,,
\label{Ddef}
\end{equation}
or alternatively 
\begin{equation}
\Delta(x,y;A) (-i\,
\buildrel{\leftarrow}\over{{D\!\!\!\!/}}_y + m) \equiv S(x,y;A)\,,
\label{Ddef2}
\end{equation}
then the function $\Delta$ satisfies the equation:
\begin{equation}
({D\!\!\!\!/}_x^{~2}+m^2)\Delta(x,y;A) = - \delta^4(x-y)\,,
\nonumber
\end{equation}
which can be written after some algebraic manipulations as
\begin{equation}
\left( -D_x^2-m^2 + {1\over 2} g \,\sigma^{\mu\nu} F_{\mu\nu}(x) \right)
\Delta(x,y;A) =  \delta^4(x-y)\,,
\label{Deq}
\end{equation}
with $F_{\mu\nu} \equiv i\left[ D_\mu,D_\nu\right]/g$ and 
$\sigma^{\mu\nu} \equiv i \left[\gamma^\mu,\gamma^\nu\right]/2$.
In what follows it is useful to introduce the operator  
$H(x,\partial_x)\equiv \left(D^2_x+m^2\right)/2 -g \,\sigma^{\mu\nu} 
F_{\mu\nu}(x)/4$. Therefore, Eq. (\ref{Deq}) can be written as 
\begin{equation}
-2~H(x,\partial_x) ~\Delta(x,y;A) = \delta^4(x-y)\,. 
\label{Deq2}
\end{equation}

Following \cite{pes83} we  consider the equation
\begin{equation}
\left( i~{d\over dT} - H(x,\partial_x) \right) 
\Phi(x,y;A;T) = i~\delta(T-T_0)~\delta^4(x-y),  
\label{Peq}
\end{equation}
with boundary condition $\Phi(x,y;A,T) = 0$ for $T<T_0$. 
The parameter $T$ is usually called proper time. 
Eq. (\ref{Peq}) is Schr\"odinger-like. The solution can be 
written as a path-integral over all the trajectories joining 
the point $y$ at time $T_0$ and the point $x$ at time $T$ 
(see e.g. \cite{sak}):
\begin{eqnarray}
\Phi(x,y;A;T) &=& \theta(T-T_0) ~ {\cal Z}(x,y;T;A) \,, \nonumber\\
{\cal Z}(x,y;T;A) &=&
\int_{y = z(T_0)}^{x = z(T)} {\cal D}z~{\cal D}p~
{\rm P}\,e^{\displaystyle i~\int_{T_0}^T dt~p\dot z - H(z,p)} \nonumber \,.
\end{eqnarray}
Integrating Eq. (\ref{Peq}) in $\displaystyle \int_{T_0}^\infty dT$ 
and taking in account that ${\cal Z}(x,y;T_0;A) = \delta^4(x-y)$, 
we obtain  the solution of Eq. (\ref{Deq2}) as 
\begin{eqnarray}
\Delta(x,y;A) &=& -{i\over 2} \int_{T_0}^\infty dT ~\Phi(x,y;A;T) \nonumber\\
&=& -{i\over 2}\int_{T_0}^\infty dT~
\int_{y = z(T_0)}^{x = z(T)} {\cal D}z~{\cal D}p~
{\rm P}\,e^{\displaystyle i~\int_{T_0}^T dt~p\dot z - H(z,p)} \,.
\label{Dpath}
\end{eqnarray}
Since the dependence on the momenta is Gaussian, the explicit integration 
on $p$ is possible. Performing it we obtain  
\begin{equation}
\Delta(x,y;A) = -{i\over 2}\int_{T_0}^\infty dT~
\int_{y = z(T_0)}^{x = z(T)} {\cal D}z~
{\rm P}\,e^{\displaystyle -  i~\int_{T_0}^T dt~  {\dot z^2 + m^2\over 2} 
- g A^\mu(z)\, \dot z_\mu - {1\over 4} g\, \sigma^{\mu\nu} F_{\mu\nu}(z)} \,.
\label{Dpath2}
\end{equation}
Eq. (\ref{Dpath2}) with (\ref{Ddef}) or (\ref{Ddef2}) supplies us with 
  a path 
integral representation of the fermion propagator in external field. 
We call this representation the Feynman--Schwinger representation of the 
fermion propagator (some historical references are given in \cite{fey}). 

\section{The 4-point Green function and the Wilson loop}

\begin{figure}[htb]
\vskip 8truecm
\makebox[4.5truecm]{\phantom b}
\epsfxsize=6truecm
\epsffile{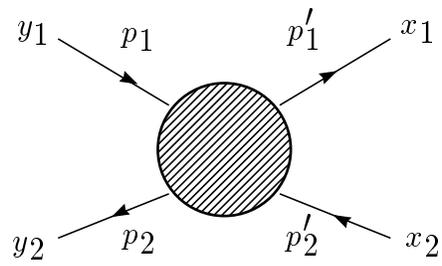}
\vskip 0.3truecm
\caption{{{\it The 4-point Green function.}}}
\label{figgreen}
\vskip 3.5truecm
\end{figure}

\begin{figure}[bht]
\vskip 0.8truecm
\makebox[3.8truecm]{\phantom b}
\epsfxsize=11truecm
\epsffile{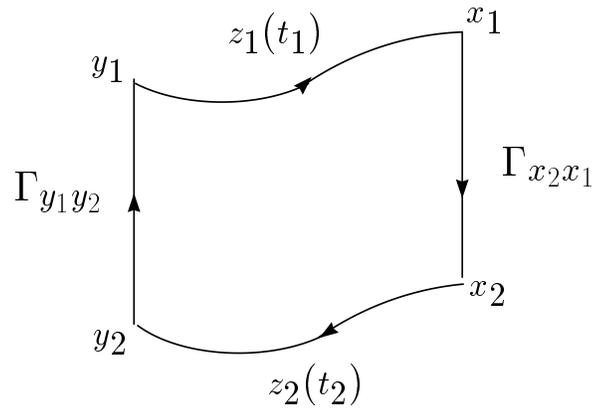}
\vskip -7truecm
\caption{{{\it The closed loop $\Gamma$.}}}
\label{figwil}
\vskip 0.8truecm
\end{figure}

Let us now consider a fermion-antifermion system. The corresponding 
4-point Green function (see Fig. \ref{figgreen}) is given by
\begin{equation}
G(x_1,x_2,y_1,y_2) = {1\over \cal N} 
\int {\cal D} \psi ~ {\cal D} \bar\psi ~ {\cal D} A ~ 
e^{\displaystyle i\int d^4x\,{\cal L}(\psi,\bar\psi,A)}
\bar\psi(x_2)\, \psi(x_1)\, \bar\psi(y_1)\, \psi(y_2),
\end{equation}
where $\cal N$ is a normalization factor and $\cal L$ is the Lagrangian 
density of the gauge theory which we are considering (in our case QCD; 
in the following  ${\cal L}_{\rm YM}$ will denote the 
Yang--Mills part of this Lagrangian density). 
Since the Lagrangian is quadratic in the fermion fields, it is possible 
to perform explicitly the integration over it. Neglecting
\begin{description}
\item{$i$)} fermion loops (quenched approximation), 
\item{$ii$)}annihilation graphs,
\end{description}
we obtain \cite{report,sitj} 
\begin{eqnarray}
G(x_1,x_2,y_1,y_2) &=&  
{1\over \cal N} \int {\cal D} A\, 
e^{\displaystyle i\int d^4x\,{\cal L}_{\rm YM}(A)} 
i\,S(x_1,y_1;A) \, i\,S(y_2,x_2;A)
\nonumber\\
&\equiv& \left\langle i\,S(x_1,y_1;A) \, i\,S(y_2,x_2;A) \right\rangle .
\label{Gfour}
\end{eqnarray}
In order to deal with gauge invariant quantities, we will consider 
in place of the above gauge dependent Green function, the so-called 
gauge invariant Green function $G_{\rm inv}$ obtained from 
the previous one by connecting the end points with the path-ordered operator 
\begin{equation}
U(y,x;\Gamma_{yx}) \equiv {\rm P}\,
e^{\displaystyle ig\int_{\Gamma_{yx}} dz^\mu A_\mu(z)} ,
\label{po}
\end{equation}
where the integration goes over an arbitrary path $\Gamma_{yx}$ 
connecting $x$ with $y$.  
Within the approximations $i)$ and $ii)$ we have 
\begin{equation}
G_{\rm inv}(x_1,x_2,y_1,y_2) =  
\left\langle {\rm Tr} \, i\,S(x_1,y_1;A) U(y_1,y_2;\Gamma_{y_1y_2}) \, 
i\,S(y_2,x_2;A) U(x_2,x_1;\Gamma_{x_2x_1})\right\rangle .
\label{Ginv}
\end{equation}
Writing now the fermion propagators in terms of the Feynman--Schwinger 
path integral representation given in the previous section, we obtain 
\begin{eqnarray}
G_{\rm inv}(x_1,x_2,y_1,y_2) &=&
{1\over 4} \Bigg\langle {\rm Tr}\,{\rm P}\, 
(i\,{D\!\!\!\!/}_{x_1}^{\,(1)}+m)\, 
\int_{T_{10}}^\infty dT_1\int_{y_1 = z_1(T_{10})}^{x_1 = z_1(T_1)}{\cal D}z_1
e^{\displaystyle - i\,\int_{T_{10}}^{T_1}dt_1 {m^2+\dot z_1^2 \over 2}   }
\nonumber\\
&\times&
\int_{T_{20}}^\infty dT_2\int_{x_2 = z_2(T_{20})}^{y_2 = z_2(T_2)}{\cal D}z_2
e^{\displaystyle - i\,\int_{T_{20}}^{T_2}dt_2 {m^2+\dot z_2^2 \over 2}   }
e^{\displaystyle ig \oint_\Gamma dz^\mu A_\mu(z)}
\nonumber\\
&\times& 
e^{\displaystyle i\,\int_{T_{10}}^{T_1}dt_1 {g\over 4}\sigma_{\mu\nu}^{(1)}
F^{\mu\nu}(z_1)}
e^{\displaystyle i\,\int_{T_{20}}^{T_2}dt_2 {g\over 4}\sigma_{\mu\nu}^{(2)}
F^{\mu\nu}(z_2)} 
(-i\,\buildrel{\leftarrow}\over{{D\!\!\!\!/}}_{x_2}^{\,(2)} + m)
\Bigg\rangle \,, \nonumber\\
\label{Ginv2}
\end{eqnarray} 
where the upper-scripts $^{(1)}$ and $^{(2)}$ refer to the first and 
second fermion line. $\Gamma$ is the closed loop defined 
by the quark trajectories $z_1 (t_1)$ and $z_2 (t_2)$ running 
from $y_1$ to $x_1$ and from $x_2$ to $y_2$ 
as $t_1$ varies from $T_{10}$ to $T_1$ and $t_2$ from $T_{20}$ to $T_2$, 
and from the paths $\Gamma_{y_1y_2}$ and $\Gamma_{x_2x_1}$ 
(see Fig. \ref{figwil}). The quantity 
\begin{equation}
W(\Gamma;A) \equiv   
{\rm Tr \,} {\rm P\,} e^{\displaystyle ig \oint_\Gamma dz^\mu A_\mu (z)},
\label{wilson}
\end{equation}
is known as the Wilson loop \cite{wilson}.

Let us make some general statements. 
The variation with respect to the path of the  path ordered operator 
$U$ is given by (see for example \cite{dura,mig}): 
\begin{eqnarray}
\delta\,U(y,x;\Gamma_{yx}) &=& 
i\,g\,{\rm P} \Bigg\{
\delta y^\mu A_\mu(y)\,U(y,x;\Gamma_{yx}) - 
\delta x^\mu A_\mu(x)\,U(y,x;\Gamma_{yx})  \nonumber\\
&~& \quad 
 -\int_0^1 ds {\dot z^\mu\delta z^\nu - \dot z^\nu \delta z^\mu \over 2} 
F_{\mu\nu}(z(s))\,U(y,x;\Gamma_{yx}) \Bigg\} ,
\label{dume}
\end{eqnarray}
where we have assumed the path $\Gamma_{xy}$ to be parameterized by the proper 
time $s$ in such a way that $z(0) = x$ and $z(1) = y$. 
From it we have immediately:
\begin{eqnarray}
{\delta\,U(y,x;\Gamma_{yx}) \over \delta S^{\mu\nu}(z)} &=& 
-i\,g\,{\rm P} \left\{ F_{\mu\nu}(z) \,U(y,x;\Gamma_{yx}) \right\} ,
\label{var1}\\
{\delta\,U(y,x;\Gamma_{yx}) \over \delta y^\mu} &=&
i\,g\,{\rm P}\left\{ A_\mu(y)\,U(y,x;\Gamma_{yx}) 
- \int_0^1ds\, \dot z^\rho  {\delta  z^\lambda \over \delta y^\mu} 
F_{\rho\lambda}(z) \,U(y,x;\Gamma_{yx}) \right\} , 
\label{var2}\\
{\delta\,U(y,x;\Gamma_{yx}) \over \delta x^\mu} &=&
i\,g\,{\rm P}\left\{ - A_\mu(x)\,U(y,x;\Gamma_{yx}) 
- \int_0^1ds\, \dot z^\rho  {\delta  z^\lambda \over \delta x^\mu} 
F_{\rho\lambda}(z) \,U(y,x;\Gamma_{yx}) \right\} , \nonumber\\ 
\label{var3}
\end{eqnarray}
where $\delta S^{\mu\nu}(z) = dz^\mu\delta z^\nu - dz^\nu \delta z^\mu$
is the infinitesimal area. 

Let us now go back to the Wilson loop (\ref{wilson}). 
As a consequence of Eq. (\ref{var1}) the insertion 
of a field strength tensor $F_{\mu\nu}$ on a point $\bar z$ of the 
loop $\Gamma$ in presence of the Wilson loop $W$ can be written as  
\begin{equation}
{\delta \, W(\Gamma,A) \over \delta \, S^{\mu\nu}(\bar z) } = 
-i\,g\,{\rm P}\left\{ W(\Gamma,A)\, F_{\mu\nu}(\bar z) \right\} . 
\label{var1w}
\end{equation}
This is known as the Mandelstam relation. 
Let us now assume that the string $\Gamma_{x_2x_1}$ is a straight line. 
This is always possible since the string  is arbitrary. 
We parameterize  $\Gamma_{x_2x_1}$ as 
$z^\mu(s) = x^\mu_1 + s(x_2 - x_1)^\mu$. 
From Eqs. (\ref{var2}) and (\ref{var3}) we have
\begin{eqnarray}
{\delta\,U(x_2,x_1;\Gamma_{x_2x_1}) \over \delta x_2^\mu} &=&
i\,g\,{\rm P}\bigg\{ A_\mu(x_2)\,U(x_2,x_1;\Gamma_{x_2x_1}) \nonumber\\
&~&\quad 
- \int_0^1ds\,s\, (x_2 - x_1)^\rho  
F_{\rho\mu}(x_1+s(x_2-x_1)) \,U(x_2,x_1;\Gamma_{x_2x_1}) \bigg\} , 
\nonumber\\
\label{var2w}\\
{\delta\,U(x_2,x_1;\Gamma_{x_2x_1}) \over \delta x_1^\mu} &=&
 i\,g\,{\rm P}\bigg\{ - A_\mu(x_1)\,U(x_2,x_1;\Gamma_{x_2x_1}) \nonumber\\ 
&~&\quad
- \int_0^1ds\,(1-s)\, (x_2 - x_1)^\rho  
F_{\rho\mu}(x_1+s(x_2-x_1)) \,U(x_2,x_1;\Gamma_{x_2x_1}) \bigg\} . 
\nonumber\\ 
\label{var3w}
\end{eqnarray}
Therefore we have 
\begin{eqnarray}
&~& \partial^\mu_{x_1} \langle {\rm Tr} \, \Delta(x_1,y_1;A) \cdots 
U(x_2,x_1;\Gamma_{x_2x_1})\rangle \nonumber\\
&=&
\langle {\rm Tr} \, \partial^\mu_{x_1} \Delta(x_1,y_1;A) \cdots 
U(x_2,x_1;\Gamma_{x_2x_1})\rangle + 
\langle {\rm Tr} \, \Delta(x_1,y_1;A) \cdots 
\partial^\mu_{x_1} U(x_2,x_1;\Gamma_{x_2x_1})\rangle \nonumber\\
&=&
\langle {\rm Tr} \, \partial^\mu_{x_1} \Delta(x_1,y_1;A) \cdots 
U(x_2,x_1;\Gamma_{x_2x_1})\rangle \nonumber\\
&+& \langle {\rm Tr\, P \,} \Delta(x_1,y_1;A) \cdots 
 U(x_2,x_1;\Gamma_{x_2x_1})\left(-i\,g\,A^\mu(x_1)\right)\rangle 
- \int_0^1 ds\,(1-s)\,(x_2-x_1)_\rho \nonumber\\ 
&~&\qquad\qquad 
\times \langle {\rm Tr\, P \,} \Delta(x_1,y_1;A) \cdots 
U(x_2,x_1;\Gamma_{x_2x_1}) \,i\,g\, F^{\rho\mu}(x_1+s(x_2-x_1))\rangle , 
\nonumber 
\end{eqnarray}
and finally (taking also in account (\ref{var1}))
\begin{eqnarray}
&~& \langle {\rm Tr} \, D^\mu_{x_1} \Delta(x_1,y_1;A) \cdots 
U(x_2,x_1;\Gamma_{x_2x_1})\rangle  = 
\nonumber\\
&~&\qquad   
\left( \partial^\mu_{x_1} - \int_0^1 ds\,(1-s)\,(x_2-x_1)_\rho 
{\delta \over \delta S_{\rho\mu}(x_1+s(x_2-x_1))}\right) \nonumber\\
&~&\qquad\qquad \times
\langle {\rm Tr} \, \Delta(x_1,y_1;A) \cdots 
U(x_2,x_1;\Gamma_{x_2x_1})\rangle . 
\label{var3cov}
\end{eqnarray}
In an analogous way we obtain 
\begin{eqnarray}
&~& \langle {\rm Tr} \cdots  \Delta(y_2,x_2;A) 
\buildrel{\leftarrow}\over{D}^\mu_{x_2}  
U(x_2,x_1;\Gamma_{x_2x_1})\rangle = 
\nonumber\\
&~&\qquad 
\langle {\rm Tr} \cdots \Delta(y_2,x_2;A) U(x_2,x_1;\Gamma_{x_2x_1})\rangle\,  
\nonumber\\
&~&\qquad\qquad \times
\left( \buildrel{\leftarrow}\over{\partial}^\mu_{x_2} - 
\int_0^1 ds\,s\,(x_2-x_1)_\rho {\buildrel{\leftarrow}\over{\delta} 
\over \delta S_{\rho\mu}(x_1+s(x_2-x_1))}\right). 
\label{var2cov}
\end{eqnarray}

Therefore Eq. (\ref{Ginv2}) can be written as 
\begin{eqnarray}
&~& G_{\rm inv}(x_1,x_2,y_1,y_2) =\nonumber\\
&~&\qquad {1\over 4}  
\left(i\,{\partial\!\!\!/}_{x_1} - i\,\gamma^\mu 
\int_0^1 ds\,(1-s)\,(x_2-x_1)^\rho 
{\delta \over \delta S^{\rho\mu}(x_1+s(x_2-x_1))}+m\right)^{(1)}
\nonumber\\
&~&\qquad\times 
\int_{T_{10}}^\infty dT_1\int_{y_1 = z_1(T_{10})}^{x_1 = z_1(T_1)}{\cal D}z_1
e^{\displaystyle - i\,\int_{T_{10}}^{T_1}dt_1 {m^2+\dot z_1^2 \over 2}   }
\nonumber\\
&~&\qquad \times
\int_{T_{20}}^\infty dT_2\int_{x_2 = z_2(T_{20})}^{y_2 = z_2(T_2)}{\cal D}z_2
e^{\displaystyle - i\,\int_{T_{20}}^{T_2}dt_2 {m^2+\dot z_2^2 \over 2}   }
\nonumber\\
&~&\qquad \times 
e^{\displaystyle - \int_{T_{10}}^{T_1}dt_1 {\sigma_{\mu\nu}^{(1)}\over 4}
{\delta \over \delta\,S_{\mu\nu}(z_1)}   }
e^{\displaystyle - \int_{T_{20}}^{T_2}dt_2 {\sigma_{\mu\nu}^{(2)}\over 4}
{\delta \over \delta\,S_{\mu\nu}(z_2)}   }
\langle W(\Gamma,A) \rangle
\nonumber\\
&~&\qquad \times 
\left( - i\,{\buildrel{\leftarrow}\over{{\partial\!\!\!/}}}_{x_2} 
+ i\,\gamma^\mu \int_0^1 ds\,s\,(x_2-x_1)^\rho {\buildrel{\leftarrow}
\over{\delta}  \over \delta S^{\rho\mu}(x_1+s(x_2-x_1))}+m\right)^{(2)}
. \label{Ginv3}
\end{eqnarray} 
All the dynamical information are contained in the Wilson loop average 
$\langle W(\Gamma,A) \rangle$ and in its functional derivatives. 
The analogous happens in potential theory where it is possible 
to express the potential up to order $1/m^2$ only in terms of 
the Wilson loop functional derivatives \cite{BMP,si,bra97}. 
If we were able to know exactly the Wilson loop average over the gauge 
fields, then we could express the 4-point quenched Green function as 
a pure quantomechanical path integral (which is very convenient also 
for numerical applications see for example \cite{nitj}). 
This would realize the Migdal program of \cite{mig}. Of course the difficult 
point is to give an evaluation of the Wilson loop.  
In the next section we will discuss the QED case, for which the Wilson 
loop average is exactly known in analytic closed form. In particular 
in order to see how Eq. (\ref{Ginv3}) works  we will show how to recover 
the binding interaction kernel in terms of Feynman graphs.  
 
\section{An exact application: QED}

Since in QED the Yang--Mills Lagrangian is quadratic 
in the fields, the Wilson loop average can be evaluated exactly.  
The result is 
\begin{equation}
\langle W(\Gamma,A) \rangle  = 
e^{ \displaystyle 
- {g^2 \over 2} \oint_\Gamma dx^\mu 
\oint_\Gamma dy^\nu  D_{\mu\nu}(x-y)},
\label{wiqed}
\end{equation}
were $D_{\mu\nu}(x-y) = \langle {\rm T} A_\mu(x) \, A_\nu(y) \rangle$ 
and $g$ has now to be interpreted as the electron electromagnetic charge.  
In the quenched approximation $D_{\mu\nu}$ is nothing else than the 
photon free propagator. Let us consider Eq. (\ref{wiqed}) 
only up to order $g^2$. Therefore our expression for the Wilson 
loop in QED will be 
\begin{equation}
\langle W(\Gamma,A) \rangle  = 
1- {g^2 \over 2} \oint_\Gamma dx^\mu \oint_\Gamma dy^\nu  D_{\mu\nu}(x-y). 
\label{wiqeda}
\end{equation}

Limiting ourselves to the Wilson loop expression (\ref{wiqeda}), 
the next step will be the evaluation of the area derivatives 
which appear in (\ref{Ginv3}). A simple calculation leads to 
\begin{equation}
{\delta \langle W(\Gamma,A) \rangle \over \delta S^{\alpha\beta}(z)}
 = -g^2 \oint_\Gamma dy^\nu \left(
\partial_\beta D_{\alpha\nu} (z-y) - \partial_\alpha D_{\beta\nu} (z-y)
\right). 
\label{wqeds1}
\end{equation}
As a particular case we have 
\begin{eqnarray}
&~&\int_0^1 ds \,(1-s)\, (x_2-x_1)^\rho
{\delta \langle W(\Gamma,A) \rangle\over \delta S^{\rho\mu}(x_1 + s(x_2-x_1))} 
= -g^2 \oint_\Gamma dy^\nu D_{\mu\nu}(x_1-y) \nonumber\\
&~&\qquad\qquad\qquad 
g^2 \oint_\Gamma dy^\nu \int^1_0 ds 
\left. \partial_\mu \left(x^\rho D_{\rho\nu}(x+x_2-y)\right)
\right|_{x = s(x_1-x_2)},
\label{wqeds2}\\
&~&\int_0^1 ds \,s\, (x_2-x_1)^\rho
{\delta \langle W(\Gamma,A) \rangle\over \delta S^{\rho\mu}(x_1 + s(x_2-x_1))} 
= g^2 \oint_\Gamma dy^\nu D_{\mu\nu}(y-x_2) \nonumber\\
&~&\qquad\qquad\qquad 
- g^2 \oint_\Gamma dy^\nu \int^1_0 ds 
\left. \partial_\mu \left(x^\rho D_{\rho\nu}(x+x_1-y)\right)
\right|_{x = s(x_2-x_1)}.
\label{wqeds3}
\end{eqnarray}
Since  the contributions in the second line of  Eqs. (\ref{wqeds2})  
and (\ref{wqeds3}) are 
exactly canceled by the action of the derivatives  
${\partial\!\!\!/}_{x_1}$ and ${\partial\!\!\!/}_{x_2}$  
on the string $\Gamma_{x_2x_1}$, 
we will assume in the following that these derivatives do not act 
on the endpoint  strings  as well as we will take into 
account only the contribution of the first lines in Eqs. (\ref{wqeds2})  
and (\ref{wqeds3}). This will simplify the display of the results.  
Putting Eqs. (\ref{wqeds1})-(\ref{wqeds3}) in (\ref{Ginv3}) we obtain
\begin{eqnarray}
&~& G_{\rm inv}(x_1,x_2,y_1,y_2) =
\nonumber\\
&~& {1\over 4} (i\,{\partial\!\!\!/}_{x_1} + m)^{(1)} \cdots 
(-i\,{\buildrel{\leftarrow}\over{{\partial\!\!\!/}}}_{x_2} + m)^{(2)} 
\nonumber\\
&~& - {g^2\over 4}  
(i\,{\partial\!\!\!/}_{x_1} + m)^{(1)} \cdots 
\Bigg\{ {1\over 2} \oint_\Gamma dx^\mu  \oint_\Gamma dy^\nu 
D_{\mu\nu}(x-y) 
\nonumber\\
&~&\qquad 
-{i\over 4} \int_{T_{10}}^{T_1}dt_1 \oint_\Gamma dy^\nu
[\gamma^\mu,{\partial\!\!\!/}_{z_1}]^{(1)} D_{\mu\nu}(z_1-y)
-{i\over 4} \oint_\Gamma dx^\mu \int_{T_{20}}^{T_2} dt_2 
[\gamma^\nu,{\partial\!\!\!/}_{z_2}]^{(2)} D_{\mu\nu}(x-z_2)
\nonumber\\
&~&\qquad
-{1\over 16} \int_{T_{10}}^{T_1}dt_1  
\int_{T_{20}}^{T_2}dt_2  [\gamma^\mu,{\partial\!\!\!/}_{z_1}]^{(1)} 
[\gamma^\nu,{\partial\!\!\!/}_{z_2}]^{(2)} D_{\mu\nu}(z_1-z_2)
\nonumber \\
&~&\qquad
-{1\over 32}
\int_{T_{10}}^{T_1}dt_1  
\int_{T_{10}}^{T_1}dt_1^\prime
[\gamma^\mu,{\partial\!\!\!/}_{z_1}]^{(1)} 
[\gamma^\nu,{\partial\!\!\!/}_{z_1^\prime}]^{(1)} D_{\mu\nu}(z_1-z_1^\prime)
\nonumber \\
&~&\qquad
-{1\over 32}
\int_{T_{20}}^{T_2}dt_2  
\int_{T_{20}}^{T_2}dt_2^\prime
  [\gamma^\mu,{\partial\!\!\!/}_{z_2}]^{(2)} 
[\gamma^\nu,{\partial\!\!\!/}_{z_2^\prime}]^{(2)} D_{\mu\nu}(z_2-z_2^\prime)
\Bigg\} 
( - i\,{\buildrel{\leftarrow}\over{{\partial\!\!\!/}}}_{x_2} + m)^{(2)} 
\nonumber\\
&~& -{g^2\over 4} \cdots \Bigg\{ 
-i{\gamma^{\mu}}^{(1)}\oint_\Gamma dy^\nu D_{\mu\nu}(x_1-y) 
- {1\over 4} {\gamma^{\mu}}^{(1)} \int_{T_{20}}^{T_2} dt_2 
[\gamma^\nu,{\partial\!\!\!/}_{z_2}]^{(2)} D_{\mu\nu}(x_1-z_2)
\nonumber \\
&~&\qquad  
- {1\over 4} {\gamma^{\mu}}^{(1)} \int_{T_{10}}^{T_1} dt_1 
[\gamma^\nu,{\partial\!\!\!/}_{z_1}]^{(1)} D_{\mu\nu}(x_1-z_1)
\Bigg\}
( - i\,{\buildrel{\leftarrow}\over{{\partial\!\!\!/}}}_{x_2} + m)^{(2)} 
\nonumber\\
&~& - {g^2\over 4}(i\,{\partial\!\!\!/}_{x_1} + m)^{(1)} 
\cdots \Bigg\{ -i {\gamma^\nu}^{(2)} \oint_\Gamma dx^\mu D_{\mu\nu}(x-x_2) 
\nonumber\\&~&\qquad 
-{1\over 4} {\gamma^\nu}^{(2)} \int_{T_{10}}^{T_1} dt_1 
[\gamma^\mu,{\partial\!\!\!/}_{z_1}]^{(1)} D_{\mu\nu}(z_1-x_2)
-{1\over 4} {\gamma^\nu}^{(2)} \int_{T_{20}}^{T_2} dt_2 
[\gamma^\mu,{\partial\!\!\!/}_{z_2}]^{(2)} D_{\mu\nu}(z_2-x_2)
 \Bigg\}
\nonumber\\
&~& + {g^2\over 4} \cdots {\gamma^{\mu}}^{(1)} {\gamma^\nu}^{(2)} 
D_{\mu\nu}(x_1-x_2),
\label{Ginvqed}
\end{eqnarray}
where the dots indicate the path integrals and the kinematic factors 
given in (\ref{Ginv3}), precisely 
\begin{eqnarray}
\cdots \equiv 
&~& \int_{T_{10}}^\infty dT_1\int_{y_1 = z_1(T_{10})}^{x_1 = z_1(T_1)}
{\cal D}z_1
e^{\displaystyle - i\,\int_{T_{10}}^{T_1}dt_1 {m^2+\dot z_1^2 \over 2}   }
\nonumber\\
&~&\qquad \times
\int_{T_{20}}^\infty dT_2\int_{x_2 = z_2(T_{20})}^{y_2 = z_2(T_2)}{\cal D}z_2
e^{\displaystyle - i\,\int_{T_{20}}^{T_2}dt_2 {m^2+\dot z_2^2 \over 2}   }.
\label{prec}
\end{eqnarray} 

In order to recover Feynman diagrams from Eq. (\ref{Ginvqed}), 
we have to throw away the endpoints string contributions. 
Because we have already taken into account the action of the 
derivatives on these strings by neglecting the second line 
contributions in Eqs.  (\ref{wqeds2}) and  (\ref{wqeds3}), 
this can be done without any problem. Of course in this way we lose gauge 
invariance, but Feynman graphs, as usually known, are not gauge 
invariant quantities. Furthermore in QED the manifest gauge-invariance 
of the two particle state is not a relevant concept.
Moreover we will neglect self-energy contributions (which 
for usual gauges do not contribute to the binding), 
following the replacement scheme:
\begin{eqnarray}
\oint_{\Gamma} dx^\mu \oint_{\Gamma} dy^\nu &\rightarrow& 
2 \int _{y_1}^{x_1}dz_1^\mu \int _{x_2}^{y_2}dz_2^\nu , 
\nonumber\\   
\oint_{\Gamma} dy^\nu &\rightarrow& \int _{x_2}^{y_2}dz_2^\nu,
\nonumber\\  
\oint_{\Gamma} dx^\mu &\rightarrow& \int _{y_1}^{x_1}dz_1^\mu.  
\nonumber
\end{eqnarray}

Let us consider now the kinematic factors (\ref{prec}).
We define
\begin{equation}
\Delta(x-y) \equiv 
\Delta(x,y;0) = -{i\over 2}\int_{T_0}^\infty dT~
\int_{y = z(T_0)}^{x = z(T)} {\cal D}z~
e^{\displaystyle -  i~\int_{T_0}^T dt~  {\dot z^2 + m^2\over 2}} .
\label{Dfree}
\end{equation}
From the  definition  we have (see also  Eq. (\ref{Deq}))
\begin{equation}
\tilde\Delta(p) \equiv \int d^4z \, e^{ipz}\Delta(z) = 
{1\over p^2 - m^2 +i\epsilon} , 
\label{Dfou}
\end{equation} 
and 
\begin{equation}
\tilde S(p) \equiv i \int d^4z \,(i{\partial\!\!\!/}_z + m) e^{ipz}\Delta(z) = 
{i\over {p\!\!\!/} - m +i\epsilon} . 
\label{Sfou}
\end{equation}
Some properties of $\Delta$ are (see Appendix):
\begin{eqnarray} 
&i)& \int_{T_0}^\infty dT\,\int_{y = z(T_0)}^{x = z(T)} {\cal D}z~ 
e^{\displaystyle -  i~\int_{T_0}^T dt~  {\dot z^2 + m^2\over 2}}
\int_{T_0}^T dt \, f(z(t)) = 
\nonumber\\
&~&\qquad
-4\int d^4\xi\,\Delta(x-\xi)\,f(\xi)\,\Delta(\xi-y) , 
\label{pro1}\\
&ii)& \int_{T_0}^\infty dT\, \int_{y = z(T_0)}^{x = z(T)} {\cal D}z~ 
e^{\displaystyle -  i~\int_{T_0}^T dt~  {\dot z^2 + m^2\over 2}}
\int_{T_0}^T dt \, f(z(t),\dot z(t)) = 
\nonumber\\
&~&\qquad 
-4\int d^4\xi\,\int d^4\eta \, \int {d^4p\over (2\pi)^4}
e^{-ip(\xi-\eta)}
\Delta(x-\xi)\,f\left({\xi+\eta \over 2},p\right)\,\Delta(\eta-y) . 
\label{pro2}
\end{eqnarray}

Since all the calculations are more transparent in the momentum space, 
it is convenient to deal with the Fourier transform of the Green function 
$G$,  defined to be (see Fig. \ref{figgreen}):
\begin{equation}
\tilde G(p_1^\prime,p_2,p_1,p_2^\prime) \equiv
\int d^4x_1\,\int d^4x_2\,\int d^4y_1\,\int d^4y_2\,
e^{-ip_1y_1}e^{ip_2y_2}e^{ip_1^\prime x_1}e^{-ip_2^\prime x_2}
G(x_1,x_2,y_1,y_2).
\label{Gtilde}
\end{equation} 
The first term of the right-hand side of Eq. (\ref{Ginvqed}) 
is therefore nothing else than the 2-particle free propagator. 
At the order $g^2$ we have to evaluate the next four terms 
(we are considering only interaction terms). These four terms can be 
factorized out in the product of two contributions on the first fermion
line and two contributions  on the second fermion line.
Taking into account only the first fermion line and  
using Eq. (\ref{pro2}) we obtain from the first contribution a term like 
\begin{eqnarray}
&~&-2\int d^4x_1\,\int d^4y_1\, e^{ip_1^\prime x_1} e^{-ip_1 y_1}
(i{\partial\!\!\!/}_{x_1} + m)^{(1)} \int d^4\xi\, \int d^4\eta 
\,\int {d^4p\over (2\pi)^4} e^{-ip(\xi-\eta)}
\nonumber\\
&~&\qquad\qquad \times 
\Delta(x_1-\xi) \left( p^\mu  - {i\over 4} [\gamma^\mu,{\partial\!\!\!/}]
\right)^{(1)} 
\,D_{\mu\nu}\left({\xi+\eta\over 2} -  \cdots \right) \,\Delta(\eta-y_1) , 
\nonumber
\end{eqnarray}
which after a straightforward calculation ends up to be
\begin{equation}
-\gamma^{\mu \, (1)} \tilde D_{\mu\nu}(p_1^\prime-p_1 - \cdots) 
\tilde\Delta(p_1) + 
\left[ \tilde S(p_1^\prime) \gamma^\mu \tilde S(p_1) \right]^{(1)}
\tilde D_{\mu\nu}(p_1^\prime-p_1 - \cdots), 
\label{n1}
\end{equation}
where $\tilde D_{\mu\nu}$ is the Fourier transform of the 
photon propagator. The second contribution on the first line is 
\begin{equation}
\int d^4x_1\,\int d^4y_1\,  e^{ip_1^\prime x_1} e^{-ip_1 y_1}
\Delta(x_1-y_1) \gamma^{\mu \, (1)} D_{\mu\nu}(x_1 - \cdots)  = 
\gamma^{\mu \, (1)} \tilde D_{\mu\nu}(p_1^\prime-p_1 - \cdots) 
\tilde\Delta(p_1). 
\label{n2}
\end{equation}
Combining together (\ref{n1}) and (\ref{n2}) we obtain 
the usual fermion-photon vertex on the first fermion line: 
$$
\left[ \tilde S(p_1^\prime) \gamma^\mu \tilde S(p_1) \right]^{(1)}
\tilde D_{\mu\nu}(p_1^\prime-p_1 - \cdots). 
$$
Handling in the same way with the second fermion line we finally have 
\begin{eqnarray}
\tilde G(p_1^\prime,p_2,p_1,p_2^\prime) &=& 
(2\pi)^4\delta^4(p_1^\prime-p_1) \tilde S(p_1^\prime)^{(1)}
~(2\pi)^4\delta^4(p_2^\prime-p_2) \tilde S(p_2)^{(2)}
\nonumber\\
&+& 
(2\pi)^4\delta^4(p_1^\prime+p_2^\prime-p_1-p_2)
\left[ \tilde S(p_1^\prime) \,ig\gamma^\mu \tilde S(p_1) \right]^{(1)}
\left[ \tilde S(p_2) \,ig\gamma^\nu \tilde S(p_2^\prime) \right]^{(1)}
\nonumber\\
&~& \times D_{\mu\nu}(p_1^\prime-p_1).
\label{Gqed1}
\end{eqnarray}
This is the 4-point free Green function plus the one-photon 
exchange graph. We notice that in order to obtain the Bethe--Salpeter 
equation for QED all the higher order corrections  to the 
formula (\ref{wiqeda}) have to be taken in account. 
This is a little bit cumbersome (and is beyond the purposes 
of this discussion) but can be done in a systematic way with the methods 
given in \cite{bra96} for the scalar QED case. Here we want to point 
out that starting from the Wilson loop behaviour given in Eq. (\ref{wiqed}), 
and restricting ourselves to the $g^2$  contributions, we were able 
to reconstruct with this technique the ladder kernel. In a similar way, 
starting with a given confining behaviour of the QCD Wilson loop 
we should be able to reconstruct the relevant contribution 
to the quark-antiquark Bethe--Salpeter kernel.

\section{QCD confining models and conclusive remarks}

In principle the same technique can be used in order to extract 
an interaction kernel from Eq. (\ref{Ginv3}) in QCD. 
The problems arise from the fact that we do not know the exact 
analytic expression for the Wilson loop average in QCD and therefore 
we do not have the equivalent of equation (\ref{wiqed}). 
We have to resort to some model approximations and from this respect it is 
really determinant, in order to recover a meaningful result, to deal with 
a gauge-invariant quantity as the Wilson loop is. However the 
up to now available approximative expressions for the Wilson 
loop average in QCD seem to be too rough for this purpose. 
In fact in order to recover a reasonable interaction kernel from the 
Feynman--Schwinger representation of the 4-point Green function, 
the dynamics of the interaction, contained in the Wilson loop, 
cannot be anything, but it has to fit with the kinematics of the two quarks, 
represented in Eq. (\ref{Ginv3}) by the external and spin projectors 
and the kinetic energy terms inside the path integral. 
For example, this matching condition manifests itself clearly in QED 
by the cancelation of (\ref{n2}) when combined with (\ref{n1}). 
This is the difficulty  when  particles with spin are taken into account. 
It is a consequence of having expressed the field insertions  
in the Wilson loop in terms of functional derivatives on the 
Wilson loop contour. In this way not only the gluodynamics, but also 
the coupling of quarks and gluons (in the perturbative and 
non-perturbative regime) is depending from the Wilson loop behaviour.

The authors of ref. \cite{bra96} assume the Wilson loop average 
to be governed by the Wilson area law, i. e.  
\begin{equation}
\langle W(\Gamma,A) \rangle  = 
e^{ \displaystyle - \sigma \, S_{\rm min}(\Gamma)},
\label{wiarea}
\end{equation}
where $S_{\rm min}(\Gamma)$ is the minimal area enclosed 
by the closed curve $\Gamma$ and $\sigma \approx 0.2 {\rm Gev}^2$ 
is the string tension. The matching between this dynamical assumption 
and the kinematics of the quarks is successful only in the 
so-called second order formalism. In other words, only if  part of the 
kinematics is not taken into  account. 

Also more sophisticated assumptions for the Wilson loop average are 
difficult to handle in the Feynman--Schwinger framework without making 
semirelativistic approximations. In the stochastic vacuum model (SVM) 
\cite{si,svm} one assumes that 
\begin{equation}
\langle W(\Gamma,A) \rangle  = 
e^{ \displaystyle - {g^2\over 2} \int_{S(\Gamma)} dS^{\mu\nu}(u) 
 \int_{S(\Gamma)} dS^{\rho\lambda}(v) 
\langle F_{\mu\nu}(u) U(u,v;\Gamma_{uv}) 
F_{\rho\lambda}(v) U(v,u;\Gamma_{vu}) \rangle },
\label{wisvm}
\end{equation}
where the curve $\Gamma_{uv}$ connecting the points $u$ and $v$ is  
arbitrary and the integration is performed over a surface 
$S(\Gamma)$ enclosed by the curve $\Gamma$. 
For what concerns the present discussion we can neglect color indices 
(the bracket $\langle ~~\rangle$ is an identity matrix 
in colour space). In the usual straight-line parameterization of the surface 
(see for example \cite{bra97}) we introduce points belonging to 
different fermion lines evaluated at the same 
proper time which seem not to be treatable in the previous formalism. 
More convenient appears to parameterize the surface in triangle 
having a fixed vertex and two vertices running on the curve $\Gamma$. 
This corresponds essentially to choose the gauge fields in 
the so-called Fock--Schwinger gauge. We will discuss this case in 
more detail in the following. 

Let us assume that the fields $A^\mu$ satisfy the gauge condition:
\begin{equation}
(x - x_1)^\mu A_\mu(x) = 0. 
\label{Fock}
\end{equation}
As a consequence we can express $A_\mu$ in terms of the field strength 
tensor $F_{\mu\nu}$ \cite{cro80}:
$$
A_\mu(x) = \int_0^1 d\alpha\, \alpha\, (x-x_1)^\rho 
F_{\rho\mu}(x_1+\alpha(x-x_1)).
$$
This gauge is a very natural tool in the sum rules method  \cite{shifman} 
and a great deal of the existing information on non-perturbative QCD dynamics 
can be recovered working in it. 

Expanding the Wilson loop average in cumulants and taking only 
the second order ones \cite{svm}, we obtain an expression formally 
equivalent to Eq. (\ref{wiqed}) where $D_{\mu\nu}$ is now no more 
a local quantity (the gauge breaks in fact the translational 
invariance) and is defined to be 
\begin{equation}
D_{\mu\nu}(x,y) = 
\int_0^1 d\alpha\, \alpha\, (x-x_1)^\rho 
\int_0^1 d\beta\, \beta\, (y-x_1)^\sigma 
\langle F_{\rho\mu}(x_1+\alpha (x-x_1))  F_{\sigma\nu}(x_1+\beta (y-x_1)) 
\rangle .
\label{proqcd}
\end{equation}
Perturbative and non-perturbative contributions are contained 
in $\langle F_{\rho\mu} F_{\sigma\nu}\rangle$. We focus our attention 
on the non-perturbative ones which are of the type: 
\begin{equation}
\langle F_{\rho\mu}(u) F_{\sigma\nu}(v)\rangle = 
(g_{\rho\sigma} g_{\mu\nu} - g_{\rho\nu} g_{\mu\sigma})
f(a^2\,(u-v)^2),
\label{nonp}
\end{equation}
where $1/a \approx 0.2 \div 0.3$ {\rm fm} is the correlation length 
which defines the confining energy (distance) regions.  
Notice that in the limiting case $a\to 0$ the form factor 
$f$ coincides  with the gluon condensate.
In this way we are considering the Wilson loop behaviour given  by the 
stochastic vacuum model. The model \cite{si,svmrev} is based on 
the idea that the low frequency contributions in the functional integral
can be described by a simple stochastic process with a converging
cluster expansion. Assuming the existence of a finite correlation length 
$1/a$ linear confinement is obtained. The simplest formulation
is characterized by a Gaussian measure specified by the correlator 
given in Eq. (\ref{nonp}) just determined by two scales: the strength
of the correlator (the gluon condensate) and the correlation length.
This behaviour of the correlator (i. e. the exponentially 
falling off, like a gaussian, of the function $f$ as $|u-v|\gg 1/a$ 
in Euclidean space) has been directly confirmed by lattice 
calculations \cite{svmlat}.
The Wilson loop behaviour given by Eqs. (\ref{wiqed}), (\ref{proqcd}),
(\ref{nonp}) has been successfully  in applications
to the study of soft high energy scattering \cite{svmscat} as
well as to the heavy quark potential. In particular  the SVM potential
reproduces exactly static \cite{wilson}, spin-dependent \cite{eichfein}
and  velocity-dependent potentials \cite{BMP,BCP} in the appropriate 
limit \cite{bra97}. 

\begin{figure}[htb]
\vskip 0.8truecm
\makebox[2truecm]{\phantom b}
\epsfxsize=11truecm
\epsffile{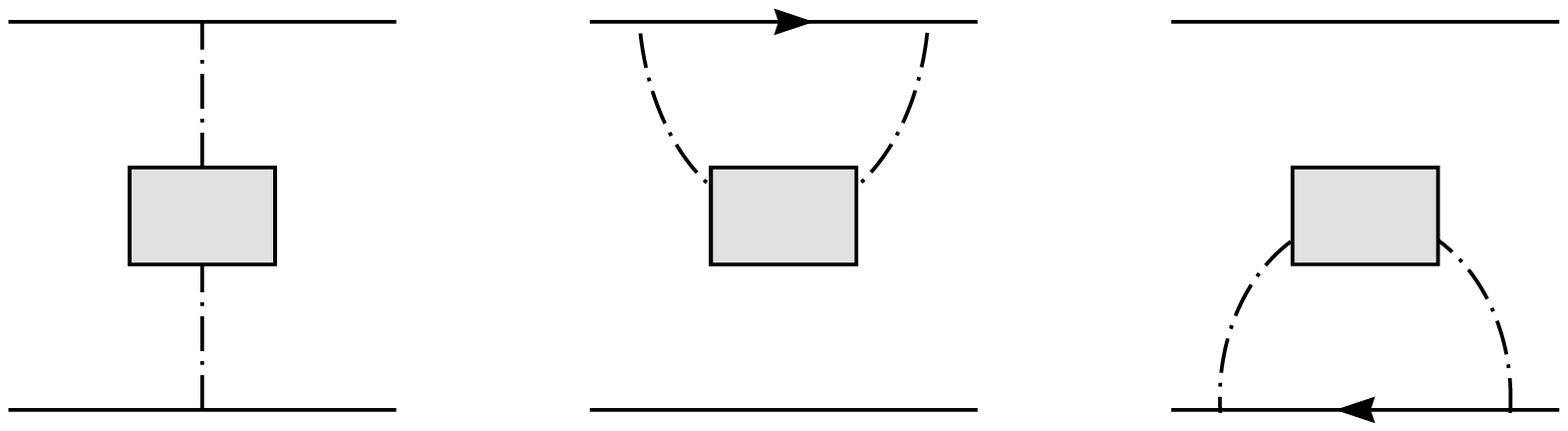}
\vskip -0.5truecm
\caption{{{\it Non-perturbative contributions to the bound state kernel 
in QCD.}}}
\label{figker}
\vskip 0.8truecm
\end{figure}

Using, now, Eqs. (\ref{wqeds1})-(\ref{wqeds3}) we obtain Eq. (\ref{Ginvqed}) 
but with the above definition of $D_{\mu\nu}$.
We observe that because of the gauge condition (\ref{Fock}),  
either the second line of Eq. (\ref{wqeds2}) and  Eq. (\ref{wqeds3}) 
vanishes  either the action of the derivatives on the string 
$\Gamma_{x_2x_1}$ is zero, being $U(x_2,x_1;\Gamma_{x_2x_1}) = 1$. 
If we assume that also in this case the endpoint string 
$U(y_1,y_2;\Gamma_{y_1y_2})$ is not relevant for the bound state 
\footnote{In this way we lose gauge invariance. Nevertheless, 
this assumption seems to be justified  by the existence of a 
finite correlation length in the correlator dynamics (\ref{nonp}).
Therefore, in the limit $(T-T_0) \to \infty $ the contribution of 
the string should be negligible. Anyway we stress that all the 
dynamics approximations have been made on gauge invariant quantities. 
Finally, in the case $y_1 = y_2$ the contribution of the 
string actually vanishes.} the Feynman graphs contributing 
to the interaction kernel are given in Fig. \ref{figker}. 
With the box we indicate the translational non invariant 
propagator $D_{\mu\nu}$ given in Eq. (\ref{proqcd}). In particular, due to  
the losing of translational invariance, the former (in QED) ``self-energy'' 
graphs, which now can be interpreted as the the interaction of 
the single quarks with the background vacuum fields,  
contribute to the binding and can not be neglected anymore. 
This last point emerges very clearly in the one body limit. 
If the mass of the particle moving on the first fermion line 
goes to infinity, then the exchange graph of Fig. \ref{figker}
does not contribute at all to the static limit which is entirely 
described by the  interaction of the second quark 
with the vacuum background fields  \cite{dir97}.
Moreover in the one body-limit we have shown \cite{dir97}  
that in this kind of graphs the confining dynamics contained 
in the correlator combines itself  with the quark propagator 
in such a way that the pole mass turns out to be  shifted by an amount $a$. 
Eventually, in the limit $a\to \infty$ the quark propagator has no pole mass. 
This means that by cutting the Feynman diagram you could not produce a 
free quark at least for some values of the parameters, which 
seems to be in the line of the results obtained by  the groups 
working  with Bethe--Salpeter and Feynman--Schwinger equations
with phenomenological kernels \cite{roberts}. 
Work is currently going on to further clarify this point.  

Finally, we observe that the Lorentz structure of the obtained kernel 
is not simply understandable in terms of a vector or scalar exchange.

\section*{Acknowledgments}

We would like to thank the Jefferson Laboratory Theory Group and the Hampton 
University for their warm hospitality as well as for the financial 
support during the first part  of this work. 
It is a pleasure to thank M. Baker and K. Maung--Maung 
for useful discussions and  strong encouragements. Discussions  
with G. M. Prosperi and C. Roberts are also acknowledged.  
The authors gratefully acknowledge the Alexander von Humboldt Foundation 
for the financial support, the perfect organization and the 
warm environment  provided to its fellows.
This work was supported in part by NATO grant under contract N. CRG960574.

\appendix
\section*{}

In this appendix we will prove Eq. (\ref{pro2}). 
Suppose to be $\epsilon$ the proper time interval in the path integral 
definition:
\begin{equation}
{\cal D}z \equiv \left( {1\over 2\pi i \epsilon} \right)^{2N-2} 
d^4z(2) \cdots d^4z(N-1). 
\end{equation}
The path integral measure satisfies the following properties
$(2<n,n+1<N-1)$: 
\begin{eqnarray}
\int_y^x {\cal D}z &=& \int d^4 z(n) \, \int_y^{z(n)} {\cal D} z  \, 
\int_{z(n)}^x {\cal D} z, \\
\int_y^x {\cal D}z &=& \int d^4 z(n) \, \int d^4 z(n+1) \, 
\left( {1\over 2\pi i \epsilon} \right)^2 
\int_y^{z(n)} {\cal D} z  \, \int_{z(n+1)}^x {\cal D} z , 
\end{eqnarray} 
and in the path integral we have 
\begin{equation}
f(z(t),\dot z(t)) \equiv  
f\left( {z(t) + z(t+\epsilon)\over 2},{z(t+\epsilon) - z(t) 
\over \epsilon} \right). 
\end{equation} 

Therefore the left-hand side of Eq. (\ref{pro2}) can be written as 
\begin{eqnarray}
&~& \int_{T_0}^\infty dT\,\int_{y = z(T_0)}^{x = z(T)} {\cal D}z~ 
e^{\displaystyle -  i~\int_{T_0}^T dt~  {\dot z^2 + m^2\over 2}}
\int_{T_0}^T dt \, f(z(t),\dot z(t)) = 
\nonumber \\ &~&\qquad~
\int d^4\xi \int d^4\eta \left( {1\over 2\pi i \epsilon} \right)^2 
\int_{T_0}^\infty dT \, \int_{T_0}^T dt 
\int_{y = z(T_0)}^{\eta = z(t)} {\cal D}z~ 
e^{\displaystyle -  i~\int_{T_0}^t d\tau ~  {\dot z^2 + m^2\over 2}}
\nonumber\\ &~&\qquad \times 
\int_{\xi = z(t+\epsilon)}^{x = z(T)} {\cal D}z~ 
e^{\displaystyle -  i~\int_{t+\epsilon}^T d\tau ~  {\dot z^2 + m^2\over 2}}
e^{\displaystyle -  i~\int_t^{t+\epsilon} d\tau ~  {\dot z^2 + m^2\over 2}}
\nonumber\\ &~&\qquad \times 
f\left( {z(t) + z(t+\epsilon)\over 2},{z(t+\epsilon) - z(t) 
\over \epsilon} \right).
\label{A1}
\end{eqnarray} 
In the  $\epsilon\to 0^+$ limit we have 
\begin{eqnarray}
&~& 
\left( {1\over 2\pi i \epsilon} \right)^2  
e^{\displaystyle -  i~\int_t^{t+\epsilon} d\tau ~  {\dot z^2 + m^2\over 2}}
f\left( {z(t) + z(t+\epsilon)\over 2},{z(t+\epsilon) - z(t) 
\over \epsilon} \right)  
\nonumber\\&~&\qquad
= \left( {1\over 2\pi i \epsilon} \right)^2 
e^{\displaystyle -i {(z(t+\epsilon) - z(t))^2 \over 2\epsilon}}
f\left( {z(t) + z(t+\epsilon)\over 2},{z(t+\epsilon) - z(t) 
\over \epsilon} \right)
\nonumber\\&~&\qquad
= f\left( {z(t) + z(t+\epsilon)\over 2}, i{d\over dz(t+\epsilon)}\right)
\left( {1\over 2\pi i \epsilon} \right)^2 
e^{\displaystyle -i {(z(t+\epsilon) - z(t))^2 \over 2\epsilon}}
\nonumber\\&~&\qquad
= f\left( {z(t) + z(t+\epsilon)\over 2}, i{d\over dz(t+\epsilon)}\right)
\int {d^4p \over (2\pi)^4} e^{-ip(z(t+\epsilon) - z(t))}
\nonumber\\&~&\qquad
= \int {d^4p \over (2\pi)^4} e^{-ip(z(t+\epsilon) - z(t))}
f\left( {z(t) + z(t+\epsilon)\over 2}, p\right).
\label{A2}
\end{eqnarray}
Putting now (\ref{A2}) in Eq. (\ref{A1}) and taking 
in account that $\displaystyle \int_{T_0}^\infty dT \int_{T_0}^T dt 
= \int_{T_0}^\infty dt \int_t^\infty dT$ we obtain Eq. (\ref{pro2}).  
As a particular case of Eq. (\ref{pro2}), if $f$ does not depend 
on $\dot z$, Eq. (\ref{pro1}) follows immediately.

\vfill
\eject 

\vskip 3truecm 

\vskip 3truecm 

\clearpage

\begin{references}

\bibitem{pes83}  M. E. Peskin, in Proceedings of the 11th SLAC Inst., 
                 SLAC Rep. n.207, 151 ed. by P. Mc. Donough (1983);
\bibitem{BMP}    A. Barchielli, E. Montaldi and G. M. Prosperi,
                 Nucl. Phys. {\bf B 296}, 625 (1988);
\bibitem{si}     Yu. A. Simonov, Nucl. Phys. {\bf B 307}, 512 (1988) 
                 and {\bf B 324}, 67 (1989);
\bibitem{report} W. Lucha, F. F. Sch\"oberl and D. Gromes, Phys. Rep. 
                 {\bf 200}, 127 (1990);
\bibitem{sitj}   Yu. A. Simonov and J. A. Tjon, Ann. Phys. {\bf 228}, 1 (1993);
\bibitem{nitj}   T. Nieuwenhuis and J. A. Tjon, Phys. Lett. {\bf B 355}, 
                 283 (1995) and  Phys. Rev. Lett. {\bf 77}, 814 (1996);
\bibitem{BCP}    N. Brambilla, P. Consoli and G. M. Prosperi, Phys. Rev.
                 {\bf D 50},  5878 (1994); N. Brambilla and G. M. Prosperi, in
                 Proceedings of ``Quark Confinement and the 
                 Hadron Spectrum'', eds. N. Brambilla and G. M. Prosperi, 
                 p. 113, (World Scientific, Singapore, 1995);
\bibitem{bra96}  N. Brambilla, E. Montaldi and G. M. Prosperi, 
                 Phys. Rev. {\bf D 54}, 3506 (1996);
\bibitem{bakbra} M. Baker, J. S. Ball, N. Brambilla, G. M. Prosperi and
                 F. Zachariasen, Phys. Rev. {\bf D 54}, 2829 (1996);
\bibitem{bra97}  N. Brambilla and A. Vairo, {\it Heavy quarkonia: Wilson 
                 area law, stochastic vacuum model and dual QCD}, 
                 Phys. Rev. {\bf D 55}, 3974 (1997); M. Baker, J. S. Ball, 
                 N. Brambilla and A. Vairo,  Phys. Lett. {\bf B 389}, 
                 577 (1996);
\bibitem{eichfein} E. Eichten and F. Feinberg, Phys. Rev. {\bf D 23}, 
                   2724 (1981);
\bibitem{sicon}    Yu. A. Simonov, {\it Confinement} (unpublished) (1996);
\bibitem{buck}     C. R. Munz, J. Resag, B. C. Metsch and 
                   H. R. Petry, Nucl. Phys. {\bf A 578}, 418  (1994);
                   A. Afanesev and W. Buck, CEBAF-TH-96-12;
\bibitem{dif} A. Gara, B. Durand and L. Durand, Phys. Rev. {\bf D 42}, 
              1651 (1990); {\bf D 40}, 843 (1989); J. F. Lagae, 
              Phys. Rev. {\bf D 45}, 305 (1992) and 317 (1992); 
              N. Brambilla and G. M. Prosperi,
              Phys. Rev. {\bf D 48}, 2360 (1993); {\bf D 46}, 1096 (1992);
              M. G. Olsson, S. Veseli and K. Williams, Phys. Rev. {\bf D 52}, 
              5141 (1995); J. Parramore and J. Piekarewicz, Nucl. Phys. 
              {\bf A 585}, 705 (1995);
\bibitem{roberts} See e.g. M. R. Frank and C. D. Roberts, Phys. Rev. 
                 {\bf C 53}, 390 (1996) and references therein;  
\bibitem{sak}     J. J. Sakurai, Modern Quantum Mechanics, ed. by 
                  The Benjamin/Cummings Publishing Company (1985);
\bibitem{fey}     R. P. Feynman, Phys. Rev. {\bf 80}, 440 (1950); 
                  J. Schwinger, Phys. Rev. {\bf 82}, 664 (1951); 
\bibitem{wilson} K. G. Wilson, Phys. Rev. {\bf D 10}, 2445 (1974);  
\bibitem{dura}   L. Durand and E. Mendel, Phys. Lett. {\bf B 85}, 241 (1979); 
\bibitem{mig}    A. A. Migdal, Phys. Rep.{\bf 102}, 199 (1983); 
\bibitem{svm}    H. G. Dosch, Phys. Lett. {\bf B 190}, 177 (1987); 
                 H. G. Dosch and Yu. A. Simonov, Phys. Lett. {\bf B 205}, 
                 339 (1988); 
\bibitem{cro80}  C. Cronstr\"om, Phys. Lett. {\bf B 90}, 267 (1980);
\bibitem{shifman}M. A. Shifman, A. I. Vainshtein and V. I. Zakharov,
                 Nucl. Phys. {\bf B 147}, 385 (1979); 
\bibitem{svmrev} H. G. Dosch, Prog. Part. Nucl. Phys. {\bf 33}, 121 (1994);
                 O. Nachtmann, Lectures given at the International 
                 University School of Nuclear and Particle Physics 
                 Schladming, Austria, (1996) HD-THEP-96-38;
\bibitem{svmlat} M. Campostrini, A. Di Giacomo and G. Mussardo,
                 Z. Phys. {\bf C 25}, 173 (1984); A. Di Giacomo and H.
                 Panagopoulos, Phys. Lett. {\bf B 285}, 133 (1992);
                 A. Di Giacomo, E. Meggiolaro and H. Panagopuolos,
                 (March 1996) hep--lat/9603017;
\bibitem{svmscat} H.G. Dosch, E. Ferreira and A. Kr\"amer, Phys. Rev. 
                  {\bf D 50}, 1992 (1994); M. R\"uter and H. G. Dosch, 
                  Phys. Lett. {\bf B 380}, 177 (1996);
\bibitem{dir97}  N. Brambilla and A. Vairo, {\it Non-perturbative 
                 dynamics of the heavy-light quark system in the 
                 non-recoil limit}, HD-THEP-97-07 (1997), to appear 
                 in Phys. Lett. {\bf B}.
\end{references}
\end{document}